# Cold bosonic atoms in optical lattices

D. Jaksch,[1,2] C. Bruder,[1,3] J. I. Cirac,[1,2] C.W. Gardiner[1,4] and P. Zoller[1,2]
[1] Institute for Theoretical Physics, University of Santa Barbara, Santa Barbara, CA 93106-4030
[2] Institut für Theoretische Physik, Universität Innsbruck, A-6020 Innsbruck, Austria
[3] Institut für Theoretische Festkörperphysik, Universität Karlsruhe, D-76128 Karlsruhe, Germany
[4] School of Chemical and Physical Sciences, Victoria University, Wellington, New Zealand
(August 31, 1998)

The dynamics of an ultracold dilute gas of bosonic atoms in an optical lattice can be described by a Bose-Hubbard model where the system parameters are controlled by laser light. We study the continuous (zero temperature) quantum phase transition from the superfluid to the Mott insulator phase induced by varying the depth of the optical potential, where the Mott insulator phase corresponds to a commensurate filling of the lattice ("optical crystal"). Examples for formation of Mott structures in optical lattices with a superimposed harmonic trap, and in optical superlattices are presented.

Optical lattices—arrays of microscopic potentials induced by the AC Stark effect of interfering laser beams–can be used to confine cold atoms [1–7]. The quantized motion of such atoms is described by the vibrational motion within an individual well and the tunneling between neighbouring wells, leading to a spectrum describable as a band structure [3]. Near-resonant optical lattices, where dissipation associated with optical pumping produces cooling, have given filling factors of about 1 atom per 10 lattice sites [1,6]. Higher filling factors will require lower temperatures, and hence will also require minimization of the optical dissipation. This can be achieved in a far-detuned optical lattice (especially with blue detuning), where photon scattering times of many minutes have been demonstrated [2]. Thus the lattice then behaves as a conservative potential, which could be loaded with a Bose condensed atomic vapor [8,9], for which present densities would correspond to tens of atoms per lattice site.

In this Letter we will study the dynamics of ultracold bosonic atoms loaded in an optical lattice. We will show that the dynamics of the bosonic atoms on the optical lattices realizes a Bose-Hubbard model (BHM) [10–16], describing the hopping of bosonic atoms between the lowest vibrational states of the optical lattice sites, the unique feature being the full control of the system's parameters by the laser parameters and configurations.

The BHM predicts phase transition from a superfluid (SF) phase to a Mott insulator (MI) at low temperatures and with increasing ratio of the onsite interaction $U$ (due to repulsion of atoms) to the tunneling matrix element $J$ [10]. In the case of optical lattices this ratio can be varied by changing the laser intensity: with increasing depth of the optical potential the atomic wave function becomes more and more localized and the onsite interaction increases, while at the same time the tunneling matrix element is reduced. In the MI phase the density (occupation number per site) is pinned at integer $n = 1, 2, \ldots$ corresponding to a commensurate filling of the lattice, and thus represents an *optical crystal* with diagonal long range order with period imposed by the laser light. The nature of the MI phase is reflected in the existence of a finite gap $U$ in the excitation spectrum.

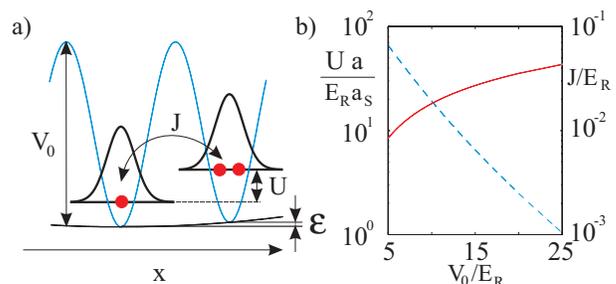

FIG. 1. a) Realization of the BHM in an optical lattice (see text). The offset of the bottoms of the wells indicates a trapping potential $V_T$. b) Plot of scaled onsite interaction $U/E_R$ multiplied by $a/a_s$ ($\gg 1$) (solid line; axis on left-hand side of graph) and $J/E_R$ (dashed line, with axis on right-hand side of graph) as a function of $V_0/E_R \equiv V_{x,y,z0}/E_R$ (3D lattice).

Our starting point is the Hamilton operator for bosonic atoms in an external trapping potential

$$H = \int d^3x \psi^\dagger(\mathbf{x}) \left( -\frac{\hbar^2}{2m}\nabla^2 + V_0(\mathbf{x}) + V_T(\mathbf{x}) \right) \psi(\mathbf{x})$$
$$+ \frac{1}{2} \frac{4\pi a_s \hbar^2}{m} \int d^3x \psi^\dagger(\mathbf{x})\psi^\dagger(\mathbf{x})\psi(\mathbf{x})\psi(\mathbf{x}) \quad (1)$$

with $\psi(\mathbf{x})$ a boson field operator for atoms in a given internal atomic state, $V_0(\mathbf{x})$ is the optical lattice potential and $V_T(\mathbf{x})$ describes an additional (slowly varying) external trapping potential, e.g. a magnetic trap (see Fig. 1a). In the simplest case, the optical lattice potential has the form $V_0(\mathbf{x}) = \sum_{j=1}^{3} V_{j0} \sin^2(kx_j)$ with wavevectors $k = 2\pi/\lambda$ and $\lambda$ the wavelength of the laser light, corresponding to a lattice period $a = \lambda/2$. $V_0$ is proportional to the dynamic atomic polarizability times the laser intensity. The interaction potential between the atoms is approximated by a short-range pseudopotential with $a_s$ the s-wave scattering length and $m$ the mass of the atoms. For single atoms the energy eigenstates are Bloch wave functions, and an appropriate superposition of Bloch states yields a set of Wannier functions which are well localized on the individual lattice sites. We assume the energies involved in the system dynamics to be small compared to excitation energies to the second band. Expanding the field operators in the Wannier basis and keeping only the lowest vibrational states, $\psi(\mathbf{x}) = \sum_i b_i w(\mathbf{x} - \mathbf{x}_i)$, Eq. (1) reduces to the Bose Hubbard Hamiltonian

$$H = -J \sum_{<i,j>} b_i^\dagger b_j + \sum_i \epsilon_i \hat{n}_i + \frac{1}{2} U \sum_i \hat{n}_i(\hat{n}_i - 1) \quad (2)$$

where the operators $\hat{n}_i = b_i^\dagger b_i$ count the number of bosonic atoms at lattice site $i$; the annihilation and creation operators $b_i$ and $b_i^\dagger$ obey the canonical commutation relations $[b_i, b_j^\dagger] = \delta_{ij}$. The parameters $U = 4\pi a_s \hbar^2 \int d^3x |w(\mathbf{x})|^4/m$ correspond to the strength of the onsite repulsion of two atoms on the lattice site $i$, $J = \int d^3x w(\mathbf{x} - \mathbf{x}_i) \left( -\frac{\hbar^2}{2m} \nabla^2 + V_0(\mathbf{x}) \right) w(\mathbf{x} - \mathbf{x}_j)$ is the hopping matrix element between adjacent sites $i, j$, and $\epsilon_i = \int d^3x V_T(\mathbf{x}) |w(\mathbf{x} - \mathbf{x}_i)|^2 \approx V_T(\mathbf{x}_i)$ describes an energy offset of each lattice site.

For a given optical potential $J$ and $U$ are readily evaluated numerically. For the optical potential given above the Wannier functions can be written as products $w(\mathbf{x}) = w(x)w(y)w(z)$ which can be determined from a one-dimensional bandstructure calculation. Figure 1b shows $U$ and $J$ as a function of $V_0$ in units of the recoil energy $E_R = \hbar^2 k^2/2m$. Both the next-nearest neighbor amplitudes and the nearest-neighbor repulsion are typically two orders of magnitude smaller and can thus be neglected. Qualitative insight into the dependence of these parameters is obtained in a harmonic approximation expanding around the minima of the potential wells. The oscillation frequencies in the wells are $\nu_j = \sqrt{4E_R V_{j0}}/\hbar$ which gives the separation to the first excited Bloch band. The oscillator ground state wave function of size $a_{j0} = \sqrt{\hbar/m\nu_j}$ allows us to obtain an estimate for the onsite interaction $U = \hbar \bar{\nu} (a_s/\bar{a}_0)/\sqrt{2\pi}$ with the bar indicating geometric means. Consistency of our model requires $a_s \ll a_{j0} \ll \lambda/2$ and $\Delta E_i = \frac{1}{2} U n_i(n_i - 1) \ll \hbar \nu_j$. The first set of inequalities follows from the pseudopotential approximation, and our requirement of a (large) energy separation from the first excited band. The second inequality expresses the requirement that the onsite interaction associated with the presence of $n_i$ particles at site $i$, which in our model is calculated in perturbation theory, must be much smaller than the excitation energy to the next band. These inequalities are readily satisfied in practice.

According to mean-field theory (MFT) in the homogeneous case [10,11] (see also [14]) the critical value of the MI - SF transition for the phase $n = 1$ is at the critical value $U/zJ \approx 5.8$ with $z = 2d$ the number of nearest neighbors. According to Fig. 1b this parameter regime is accessible by varying $V_0$ in the regime of a few tens of recoil energies. As an example, for Sodium [9] we have $E_R = 2\pi \times 8.9$ kHz for a red detuned laser with $\lambda = 985$ nm, and the critical values for the first MI phase in 1D, 2D and 3D are given by $V_{x0} = 10.8$, $V_{x,y0} = 14.4$, and $V_{x,y,z0} = 16.5 E_R$, and we assumed in 1D $V_{y,z0} = 25 E_R$ for the $y$ and $z$ directions in order to suppress tunneling in these other dimensions, and $V_{z0} = 25 E_R$ for 2D. For $V_0 = 15$ we have $U = 0.15$ and $J = 0.07$ in units of $E_R$. For a blue detuning [9] $\lambda = 514$ nm we find $E_R = 2\pi \times 32$ kHz and the corresponding values are $V_{x0} = 8.4$, $V_{x,y0} = 11.9$ and $V_0 = 14.1$, and $U = 0.2$, $J = 0.02$ for $V_0 = 10$ in units of $E_R$. For $V_0 \approx 10 E_R$ the single particle density at the center of the optical potential wells will be of order $1/a_0^3 \approx 10^{15}$ cm$^{-3}$. Thus we must discuss the role of collisions between ground state atoms (in the presence of a laser field) as a loss and decoherence mechanism [17]. This question is directly related to the problem of collisional loss of Bose-Einstein condensates in optical traps as studied in [9]. We emphasize that in the Mott phase with a single particle per site ($n = 1$) two and more particle loss channels are absent. For a MI phase with $n = 2$ there will be two particle losses: if we take as an order of magnitude the numbers published in Ref. [18] we estimate the correspondig life time to be $> 10$ s. For $n = 3$ the life time due to three atom losses [18] will be of the order $1/10$ s.

We have performed mean-field calculations for 1D and 2D configurations, as well as an exact diagonalization of the BH Hamiltonian in 1D to illustrate the formation of the Mott insulator phase in optical lattices, in particular for the inhomogeneous case. Our mean field calculations are based on a Gutzwiller ansatz for the ground state wave function $|\Psi_{MF}\rangle = \prod_i |\phi_i\rangle$ with $|\phi_i\rangle = \sum_{n=0}^\infty f_n^{(i)} |n\rangle_i$ where $|n\rangle_i$ denotes the Fock state with $n$ atoms at site $i$ [11]. We minimize the expectation value of the Hamiltonian,

$$\langle \Psi_{MF} | H | \Psi_{MF} \rangle - \mu \langle \Psi_{MF} | \sum_i \hat{n}_i | \Psi_{MF} \rangle \to \min, \quad (3)$$

with respect to the coefficients $f_n^{(i)}$. The Lagrange multiplier $\mu$ enforces a given mean particle number $N = \sum_i \langle \hat{n}_i \rangle$. This corresponds to a calculation in the grand canonical ensemble with chemical potential $\mu$ at temperature $T = 0$. A MI phase is indicated by solutions in the form of single Fock states, $|\phi_i\rangle \to |n_i\rangle_i$. A signature of a MI phase is integer occupation number (density) $\rho_i = \langle \hat{n}_i \rangle$ and fluctuations, $\sigma_i^2 = (\langle \hat{n}_i^2 \rangle - \langle \hat{n}_i \rangle^2)/\langle \hat{n}_i \rangle \to 0$. Solutions in the form of superposition of Fock states result in a mean-field $\phi_i = \langle b_i \rangle \neq 0$, indicating the presence of a SF component. The angular brackets indicate an average in the meanfield state. In the homogeneous case ($\epsilon_i = 0$) the phase diagram in the $J - \mu$ plane consists of a series of lobes [10]. Inside the lobes (i.e. for $J$ small in comparison with the onsite repulsion energy $U$) the system is a Mott phase; outside it is superfluid.

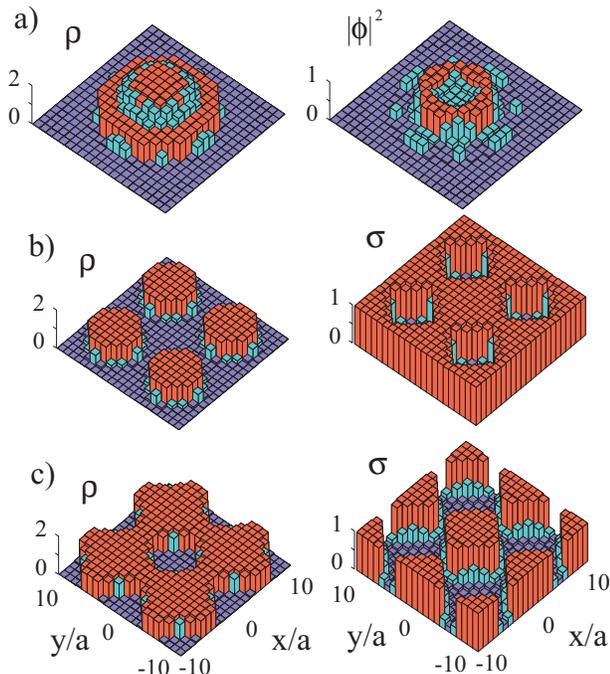

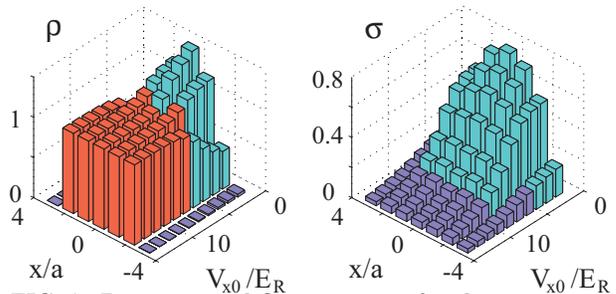

FIG. 2. a) MI and SF phases in an optical potential and harmonic trap in 2D. Parameters: $U = 35J$, $V_T(x,y) = J(x^2+y^2)/a^2$, and $\mu = 50J$. Density $\rho(x,y)$ (left plot), and superfluid density $|\phi(x,y)|^2$ (right plot). b) Superlattice in 2D. Density $\rho(x,y)$ (left plot) and fluctuations $\sigma(x,y)$ (right plot). Parameters: $U = 45J$, $V_T(x,y) = 30J\,(\sin^2(\pi x/11a) + \sin^2(\pi y/11a))$, and $\mu = 25J$. c) Same as b) with $\mu = 35J$. Four superlattice wells are shown.

In Fig. 2a we plot the density $\rho(x,y)$ and the superfluid component $|\phi(x,y)|^2$ in an optical lattice with a superimposed isotropic harmonic potential at the lattice points $(x/a, y/a) = (i,j)$ $(i,j = 0, \pm 1, \ldots)$. Fig. 2a shows a MI phase with two atoms per site at the center of the trap ($\rho = 2$) surrounded by a Mott phase with a single atom ($\rho = 1$), and superfluid rings between the MI phases. For smaller values of the chemical potential only a single Mott phase would exist at the trap center. Qualitatively, this behavior is readily understood on the basis of the phase diagram in the homogeneous case [10] if we note that the offset $\epsilon_i = V_T(\mathbf{x}_i)$ leads to an effective local chemical potential $\mu - \epsilon_i$.

By use of interfering laser beams at different angles [4], one can produce a *superlattice*, in which the offset of the optical potential is modulated periodically in space on a scale larger than the lattice period. Fig. 2b,2c, show the density $\rho(x,y)$ and the scaled density fluctuations $\sigma(x,y)$ of Mott structures formed in a superlattice. With increasing $\mu$ we first find a Mott structure at the bottom of the superlattice potential, until the atoms are no longer confined to a particular well of the superlattice but form bridges connecting the superlattice wells.

In general, specific Mott structures can be designed by an appropriate choice of the laser configurations. An experimental signature to detect the Mott state is observation of reduced density-density fluctuations (see $\sigma(x,y)$ in Fig. 2). This can be monitored directly in light scattering. Alternatively, the MI phase can be detected spectroscopically by observing the gapped particle-hole excitations.

FIG. 3. Density $\rho$ and fluctuations $\sigma$ for the exact ground state in 1D for $N = 5$ atoms in a harmonic well as a function of $V_{x0}/E_R$ for seven lattice cells. The parameters are $a_s/a = 1.1 \; 10^{-2}$ (corresponding to Na and $\lambda = 514$ nm, $V_{y0} = V_{z0} = 40E_R$) and $V_T(x) = 0.06\, E_R(x/a)^2$.

In 1D and for systems with few atoms per superlattice well we expect fluctuations to be important, and the application of MFT becomes questionable. On the other hand, in this limit it is straightforward to diagonalize the Bose Hubbard Hamiltonian exactly. Fig. 3 is a plot of the density and the number fluctuations for the exact ground state for $N = 5$ atoms as a function of $V_{x0}$. With increasing $V_{x0}$ the density shows a clear transition to the MI phase $\rho = 1$, even for this very small sample. The number fluctuations are suppressed in the MI phase, but remain finite. The phase transition (which according to MFT in the homogeneous limit is expected for $V_0 = 7.4E_R$) is smeared out, and fluctuations are strongly suppressed only for larger values of $V_{x0}$. Qualitatively, the mean field theory for the inhomogeneous case agrees well with the exact calculations, even for these small systems. Fig. 3 can be viewed as an adiabatic transfer into the MI phase as the laser intensity is varied slowly as a function of time.

The atomic level scheme of Fig. 1 allows only one adjustable parameter, the depth of the optical potential $V_0$. To adjust the tunneling matrix element $J$ independently of the onsite interaction $U$ we can employ atomic configurations with two internal ground state levels $|g_1\rangle$ and $|g_2\rangle$, which are connected by an off-resonant Raman transition (Fig. 4a).

We assume that the two internal states move in optical potentials which are shifted relative to each other by $\lambda/4$, as is the case when they have polarizabilities of opposite sign. Expanding the bosonic field operators for the two internal states we obtain a two-species Bose Hubbard Hamiltonian

$$H = -(J \sum_{<i,j>} a_i^\dagger b_j + h.\ c.) + \sum_i \epsilon_i a_i^\dagger a_i + \sum_j (\epsilon_j - \delta) b_j^\dagger b_j$$
$$+ \frac{U_{aa}}{2} \sum_i a_i^{\dagger 2} a_i^2 + U_{ab} \sum_{<i,j>} a_i^\dagger a_i b_j^\dagger b_j + \frac{U_{bb}}{2} \sum_j b_j^{\dagger 2} b_j^2 \quad (4)$$

with $a_i$ and $b_i$ bosonic destruction operators referring to atoms in the internal states $|g_1\rangle$, and $|g_2\rangle$, respectively. The first term in the Hamiltonian describes the Raman induced hopping between adjacent cells with coupling $J = \frac{1}{2} \int d^3\mathbf{x}\, w_a(\mathbf{x})^* \Omega_{\text{eff}}(\mathbf{x}) w_b(\mathbf{x} - \lambda/4)$, where $\Omega_{\text{eff}}$ is the effective two-photon Rabi frequency (including a possible phase). Direct tunneling has been neglected. The second and third term contain offsets due to a trapping potential, and in addition a Raman detuning term $-\delta$ for atoms in state $|g_2\rangle$. The second line contains onsite interactions of atoms $a$ and $b$ described by $U_{aa}$ and $U_{bb}$, and a nearest-neighbor interaction $U_{ab}$ whose value depends on the overlap of the Wannier functions between $a$ and $b$. A Raman detuning $\delta$ shifts the chemical potential of species $b$ relative to $a$. We can adjust the value of this detuning to generate checkerboard patterns, e.g. a MI phase of species a and a Mott phase of species b can coexist with different site occupation numbers. As an example, Fig. 4b plots the density $\rho(x,y)$ for a specific 2D homogeneous situation where a MI phase $|g_1\rangle$ coexists with a superfluid component in $|g_2\rangle$.

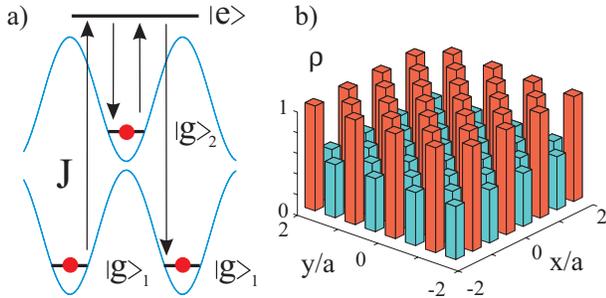

FIG. 4. a) Atomic level scheme (see text). b) Checkerboard pattern with a MI phase on one sublattice, and a SF on the other obtained in MFT for the two species BHM, with parameters: $\mu = 25J$, $U_{aa} = U_{bb} = 45J$, $U_{ab} = 0$, $\delta = -25J$, and $\epsilon_i = 0$.

While the present discussion has emphasized periodic (ordered) Bose systems, adding a further optical potential with incommensurate lattice spacing allows the realization of a (pseudo)random potential [5] which leads to the study of disordered Bose systems and appearance of a Bose glass phase [10,15]. A study of the growth and fluctuations of the MI phase due to coupling to a finite temperature particle reservoir based on a master equation treatment [19] will be presented elsewhere. The ability to manipulate both the lattice and the system parameters in our realization of a Bose-Hubbard model brings a new aspect to condensed matter physics: models and simplifying assumptions may be systematically investigated using the experimental techniques of quantum optics.

The authors thank the members of the BEC98 program at ITP UCSB for discussions. Work supported in part by the NSF, the Austrian Science Foundation, EU TMR networks and the Marsden contract PVT-603.